\documentclass[twocolumn]{jpsj2} 
\title{Dependence of Modulation Amplitude on Electron Density in Unidirectional Lateral Superlattices: The Effect of the Thickness of the Two-dimensional Electron Gas}

\author{Akira \textsc{Endo}\thanks{E-mail address: akrendo@issp.u-tokyo.ac.jp} and Yasuhiro \textsc{Iye}}

\inst{Institute for Solid State Physics, University of Tokyo, 5-1-5 Kashiwanoha, Kashiwa, Chiba 277-8581}

\abst{The amplitude $V_0$ of unidirectional periodic potential modulation introduced by a surface grating into a two-dimensional electron gas (2DEG) formed at AlGaAs/GaAs heterointerface is measured as a function of electron density $n_e$ by analyzing commensurability oscillation of the magnetoresistance. The electron density is varied either by applying a bias to a metallic back gate or by illumination. The amplitude decreases with increasing density, with the rate $|dV_0/dn_e|$ roughly an order of magnitude larger for the former method. The result is interpreted in terms of the rate, $dE_1/d(\delta E_c)$, of the change in the first subband level $E_1$ in response to the variation of the conduction-band edge $\delta E_c$ above the heterointerface. The rate crucially depends on the thickness of the 2DEG\@.}

\kword{two-dimensional electron gas, AlGaAs/GaAs single heterostructure, potential modulation, lateral superlattice, electron density, thickness, back gate, illumination, commensurability oscillation}

\begin{document}
\maketitle

\section{Introduction} 
A two-dimensional electron gas (2DEG) formed at Al$_x$Ga$_{1-x}$As/GaAs heterointerface has been the basis of a multitude of experimental studies for low-dimensional electron systems. The devices for such experiments are produced by processing the surface of the 2DEG wafers by micro- (or nano-) fabrication techniques: depositing patterned gates or etching off part of the wafers, thereby introducing electrostatic potential modulation or confining electrons to smaller areas. \cite{BeenakkerR91} The modern nano-fabrication technology grants fine control of the lateral dimensions of the patterns down to the length scale less than 100 nm, limited only by the depth of 2DEG plane from the surface. In contrast to the high controllability of the lateral length scale, the energy scale involved, the magnitude of the potential modulation, allows much poorer management. The amplitudes are often adjusted, chiefly on empirical basis, only for a certain particular purpose, e.g., depleting electrons underneath the metallic gate by applying negative bias. It is usually not easy even to measure, much less to precisely adjust, the amplitude of the modulation for generic cases. An important exception is when the modulation possesses a unidirectional periodicity. The \textit{unidirectional lateral superlattice} (ULSL) reveals the amplitude $V_0$ of the modulation through the amplitude of the commensurability oscillation (CO), the magnetoresistance oscillation resulting from the commensurability between the period $a$ of the modulation and the cyclotron radius $R_c$. \cite{Weiss89}

In the present paper, we report the behavior of the modulation amplitude when the electron density $n_e$ is varied, in pursuit of better quantitative understanding of the energy scale relevant in the study of low-dimensional electron systems. The amplitude $V_0$ is measured using CO of ULSL\@. We shed light on the effect of finite thickness of the 2DEG wave function, which is often neglected in the first order approximation but nevertheless can be of vital importance for phenomena in which electron-electron interaction plays a crucial role. \cite{FQHE}
In a ULSL, a grating that introduces potential modulation to the 2DEG plane is placed on the front surface. Therefore, it is clear that a metallic uniform front gate, a standard device to vary $n_e$, cannot be employed without heavily affecting the modulation amplitude. We have instead made use of metallic gate on the backside of the wafer, or illumination by light emitting diode (LED). Naively, one would expect the modulation amplitude to be basically insensitive to the back gate, since the grating and the back gate, being situated on the opposite side of the 2DEG, is expected to cause minimal crosstalk. It turns out, however, that the amplitude varies substantially in response to the back gate, much more pronounced than when $n_e$ is modified by illumination. The large variation is ascribed to the modification in the width of the well confining the electrons in the vertical direction, hence to the thickness of the 2DEG wave function. The experimental methods and results are described in \S \ref{Experimental} and \ref{Results}, respectively. The interpretation of the experimental results is discussed in detail in \S \ref{Discussion}, followed by concluding remarks in \S \ref{Conclusions}.

\section{Experimental}
\label{Experimental}
\begin{figure}[tb]
\includegraphics[bbllx=20,bblly=50,bburx=600,bbury=750,width=8.5cm]{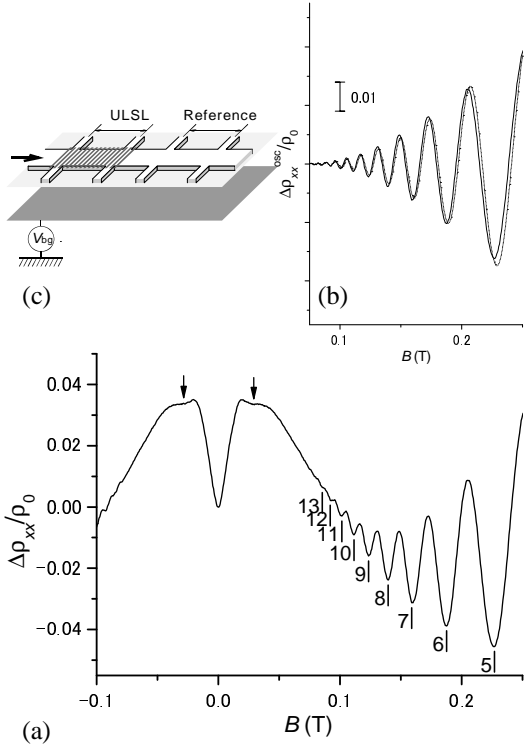}
\caption{(a) A typical magnetoresistance trace showing PMR and CO (taken at 4.2 K). (The example is for sample B at $n_e$=2.2$\times$10$^{15}$ m$^{-2}$). Arrows mark the position of $B_\mathrm{inf}$. Minima in CO are labeled by their index numbers. (b) Oscillatory part of CO, obtained by subtracting the slowly varying background. The dotted curve represents a fit to eq.\ (\ref{COAE}). (c) Schematic drawing of the sample.
\label{magres}}
\end{figure}

Two ULSL samples with differing periods were prepared from the same Al$_x$Ga$_{1-x}$As/GaAs ($x$=0.33) single-heterostructure (SH) 2DEG wafer having the mobility $\mu$$\simeq$70 m$^2$/Vs and electron density $n_e$$\simeq$2$\times$10$^{15}$ m$^{-2}$ at $T$=4.2 K\@. The structure of the wafer was (from the front surface) 10 nm GaAs cap layer, 40 nm Si-doped ($N_\mathrm{Si}$=2$\times$10$^{24}$ m$^{-3}$) Al$_x$Ga$_{1-x}$As layer, 40 nm undoped Al$_x$Ga$_{1-x}$As spacer layer, and 1 $\mu$m GaAs layer with 2DEG channel residing near the interface to the upper layer.  As depicted in Fig.\ \ref{magres}(c), a pair of Hall bars, 44$\times$16 $\mu$m$^2$, were defined in series by wet-etching; one of these was further processed into ULSL and the other was reserved as reference. The potential modulation was introduced employing strain-induced piezoelectric effect \cite{Skuras97} by a grating of high-resolution negative electron-beam resist \cite{Fujita96} placed on the front surface. \cite{Endo00e} The periods are $a$=161 and 138 nm for samples A and B, respectively. A uniform metallic (gold) gate was deposited on the rear side of the wafer after thinning the wafer from 0.3 mm down to about 0.1 mm by wet-etching in order to enhance the effectiveness of the back gate. No signs of deterioration in the mobility were observed after these processing. \cite{density} The electron density varies, by $\Delta n_\mathrm{bg}$, in response to the bias $V_\mathrm{bg}$ applied to the back gate. From the rate $d(\Delta n_\mathrm{bg})/dV_\mathrm{bg}$=5.0$\times$10$^{-3}$ and 4.1$\times$10$^{-3}$ in 10$^{15}$ m$^{-2}$/V, the distance of 2DEG plane from the back gate was estimated to be $d_\mathrm{bg}$=140 and 170 $\mu$m by a simple capacitance model $e\Delta n_\mathrm{bg}$=($\epsilon_0\epsilon/d_\mathrm{bg})V_\mathrm{bg}$ for samples A and B, respectively.

Figure \ \ref{magres}(a) shows a typical magnetoresistance trace of our ULSL, taken by standard low-frequency ac lock-in technique at 4.2 K\@. The trace exhibits CO as well as positive magnetoresistance \cite{Beton90P} (PMR) emanating from $B$=0. The amplitude $V_0$ of the unidirectional periodic potential modulation \cite{harmonics} $V_\mathrm{mod}(x)=V_0\cos(qx)$ with $q=2\pi/a$, can be deduced from the amplitude of CO\@. In order to obtain accurate quantitative value, however, care should be taken of the impurity scattering that scatters electrons out of the cyclotron orbit before completing a cycle. Such scattering reduces the CO amplitude, more effectively for smaller $B$ where cyclotron circumference 2$\pi$$R_c$ is large, and can lead to the underestimation of $V_0$ if the standard prescription \cite{Beenakker89,Peeters92} is used for the analysis. The present authors have shown \cite{Endo00e} that the scattering can be well accounted for by the inclusion of an additional factor $A(\pi/\mu_W B)$ in the formula as,
\begin{eqnarray}
\frac{\Delta\rho_{xx}^\mathrm{osc}}{\rho_0}
&\!\!\!\!\!\!\!\!=&\!\!\!\!\!\!\!\!A\left(\frac{\pi}{\mu_\mathrm{W}B}\right)A\left(\frac{T}{T_a}\right) \times \nonumber \\
& &\!\!\!\!\!\!\!\!\!\!\!\!\!\!\frac{1}{2\sqrt{2\pi}}\frac{1}{\Phi_0{\mu_\mathrm{B}^*}^2}\frac{\mu^2}{a}\frac{V_0^2}{n_e^{3/2}}|B| \sin\left(2\pi \frac{2R_\mathrm{c}}{a}\right),\label{COosc}
\label{COAE}
\end{eqnarray}
where $A(x)$$=$$x/\sinh(x)$, $k_BT_a$$=$$(1/2\pi^2)(ak_F/2)\hbar\omega_c$ with $\omega_c$$=$$eB/m^*$, $\Phi_0$=$h/e$, $\mu_\mathrm{B}^*$$\equiv$$e\hbar/2m^*$, and $R_c$$=$$\hbar k_F/e|B|$ with $k_F$$=$$(2\pi n_e)^{1/2}$ the Fermi wave number. As shown in Fig.\ \ref{magres}(b), the oscillatory part of the magnetoresistance can be fitted, with $\mu_W$ and $V_0$ as fitting parameters, to eq.\ (\ref{COAE}) very well. The parameter $\mu_W$ describes the degree of decay of CO amplitude with decreasing $B$, and was shown to be able to be identified with \cite{Endo00e} the single-particle or quantum mobility $\mu_Q$ deduced from the analysis of the decay of Shubnikov-de Haas oscillation \cite{Coleridge91} in the adjacent reference plain Hall bar. An alternative way to deduced $V_0$ is provided by the analysis of PMR\@. The present authors have recently pointed out \cite{Endo05P} that inflection points $B_\mathrm{inf}$ at which curvature of the magnetoresistance changes from concave down to concave up, as pointed by the arrows in Fig.\ \ref{magres}(a), can be found on the PMR when $V_0$ is small enough, and $B_\mathrm{inf}$ corresponds to the extinction field,
\begin{equation}
 B_\mathrm{e}=\frac{2\pi m^*V_0}{ae\hbar k_\mathrm{F}},
\label{PMRAE}
\end{equation}
where streaming orbits, the orbit of electron confined in a single period, cease to exist. The values of $V_0$ deduced by this second method agree very well with those drawn using eq.\ (\ref{COAE}), confirming that reliable values are obtained.

\section{Results}
\label{Results}
\begin{figure}[tb]
\includegraphics[bbllx=20,bblly=90,bburx=550,bbury=800,width=8.5cm]{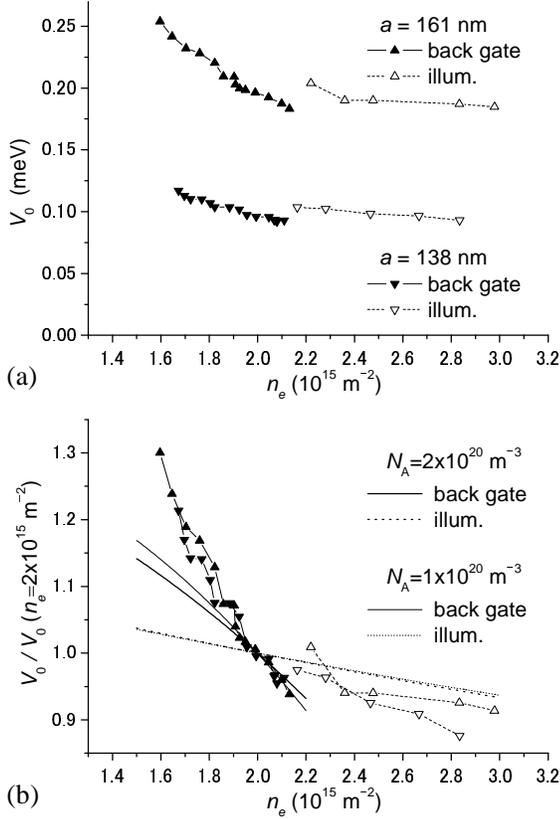}
\caption{(a) Amplitudes $V_0$ of potential modulation, deduced from CO at 4.2K using eq.\ (\ref{COAE}), plotted as a function of the electron density $n_e$. Solid and open symbols indicate that the densities were varied by the back gate or by illumination, respectively. Upward and downward triangles represent samples A ($a$=161 nm) and B ($a$=138 nm), respectively. (b) Amplitudes normalized by interpolated ($n_e$ back-gate controlled) or extrapolated ($n_e$ illumination controlled) values at $n_e$$=$$2.0\times10^{15}$ m$^{-2}$. Calculated curves simulating the variation of $n_e$ by the back gate and illumination are shown by solid and dotted curves, respectively, for $N_A$$=$1.0 and 2.0$\times10^{20}$ m$^{-3}$ as thin and thick curves, respectively. (See text for details.)
\label{V0plot}}
\end{figure}

Figure \ref{V0plot}(a) shows $V_0$ obtained using eq.\ (\ref{COAE}) from the amplitude of CO for samples A and B, plotted as a function of $n_e$. The electron density $n_e$ is varied either by back-gate voltages (solid symbols) or by illumination with an infrared LED resorting to the persistent photoconductivity effect (open symbols). The ratio of $V_0$ to the Fermi energy $E_F$ ranges from 2 to 4\% for sample A, and 1 to 2\% for sample B. It can clearly be seen that $V_0$ decreases with increasing $n_e$ for both samples. The rate of the decrease $dV_0/dn_e$ is nearly an order of magnitude larger when $n_e$ is varied by the back gate. Although $V_0$ of sample A is much larger than that of sample B, \cite{adep} the relative change, $V_0$ normalized by the value at $n_e$=2.0$\times$10$^{15}$ m$^{-2}$, shows almost the same behavior for both samples, as demonstrated in Fig.\ \ref{V0plot}(b).

Decrease of the modulation amplitude $V_0$ in response to the increase in $n_e$ by a back gate was also reported by Soibel \textit{et al}. \cite{Soibel97} As will be detailed in the next section, the decrease cannot be ascribed to the screening by 2DEG, which is independent of $n_e$ unless $a$ is less than the half of the Fermi wavelength $\lambda_F$. In the case of ref.\ \citen{Soibel97}, $V_0$, hence the ratio $V_0/E_F$, is roughly an order of magnitude larger than those of our samples. The authors of ref.\ \citen{Soibel97} pointed out that their $V_0$ is not small enough to be treated within the standard framework of the screening theory; they claimed that alteration of the 2DEG density of state by their large $V_0$ should be taken into consideration. Obviously, such explanation does not apply to our samples having $V_0$ of only a few percent of $E_F$. In the next section, we will describe our interpretation of the behavior of $V_0$ caused by the change of $n_e$.

\section{Discussion}
\label{Discussion}
\subsection{Screening}
We start by examining the effect of screening. In the Lindhard formula, the dielectric constant for an ideal zero-thickness 2DEG is given by \cite{AndoR82,DaviesB98} $\epsilon_0\epsilon\epsilon_\mathrm{s}(q)$ with $\epsilon_0$=8.85$\times$10$^{-12}$ F/m the permittivity of the vacuum, $\epsilon$ the relative dielectric constant of the host crystal ($\epsilon$=13.18 for GaAs), and
\begin{equation}
\epsilon_\mathrm{s}(q)=1+\frac{2}{a_B^*q}\left\{1-\mathrm{Re}\left[\sqrt{1-\left(\frac{2k_F}{q}\right)^2}\right]\right\},
\label{Lindhard}
\end{equation}
expressing the screening by the 2DEG\@. $a_B^*$$=$$4\pi\epsilon_0\epsilon\hbar^2/m^*e^2$ represents the effective Bohr radius for the host material (10.4 nm for GaAs). In eq.\ (\ref{Lindhard}), only the last term depends on $n_e$, which disappears at $q$$<$$2k_F$. Namely, the change in $n_e$, \textit{per se}, does not affect the screening for $a$$\geq$$\lambda_F/2$, where $\epsilon_\mathrm{s}(q)$ equals $\epsilon_\mathrm{TF}(q)$=$1+2/a_B^*q$ calculated by Thomas-Fermi approximation. For the range $n_e$=1.5--3.0$\times$10$^{15}$ m $^{-2}$ of the electron density investigated in the present study, $\lambda_F$=45--65 nm, and therefore, the screening by 2DEG of our potential modulation $V_\mathrm{mod}(x)$ will not depend on $n_e$. For realistic 2DEG with finite thickness, $\epsilon_\mathrm{TF}(q)$ should be slightly modified to include the form factor $F(q)$:
\begin{equation}
\epsilon_\mathrm{TF}(q)=1+\frac{2}{a_B^*q}F(q).
\end{equation}
The form factor is defined as \cite{Price84,Hirakawa86,Esfarjani90}
\begin{eqnarray}
F(q)&\equiv&2q\int\frac{dk}{2\pi}\frac{\tilde{f}(k)\tilde{f}(-k)}{q^2+k^2} \nonumber \\
&=&\int\!\!\!\!\int\!dzdz^\prime f(z)f(z^\prime)\exp(-q|z-z^\prime|),\label{Form}
\end{eqnarray}
using the Fourier transform $\tilde{f}(k)$ of the (normalized) electron distribution function $f(z)$$=$$|\psi(z)|^2$ in the $z$ direction, the direction normal to the 2DEG plane. $\psi(z)$ designates the envelope function of the electron wave function $\psi(z)\exp[i(k_xx+k_yy)]$. It can readily be seen that $F(q)$=1 for $f(z)$=$\delta(z)$, an ideal 2DEG\@. As a rule of thumb, $F(q)$ decreases when $f(z)$ is widely spread: the screening is less effective for thicker 2DEG\@. For example, for Fang-Howard wave function \cite{Fang66,AndoR82} $\psi_\mathrm{FH}(z)$$=$$\Theta(z)(b^3/2)^{1/2}z\exp(-bz/2)$, $F(q)$$=$$b(8b^2+9bq+3q^2)/8(b+q)^3$$=$$1-15/8(q/b)+3(q/b)^2+O[(q/b)^3]$. Here, $\Theta(z)$ represents the unit step function. The variational parameter $b$, given by $b$$=$$[48\pi(n_\mathrm{depl}+11n_e/32)/a_B^*]^{1/3}$ with $n_\mathrm{depl}$ the depletion charge, is inversely proportional to the thickness of the wave function [rms thickness $(\langle z^2\rangle-\langle z\rangle^2)^{1/2}$=$3^{1/2}/b$], and therefore $F(q)$ decreases with increasing thickness.

The thickness of 2DEG varies with $n_e$, and therefore the change in $n_e$ can, in principle, alter the modulation amplitude seen by the electrons through the change in the screening. However, it is obvious that this thickness-mediated change in the screening is unable to explain our $V_0-n_e$ relationship even in the qualitative level. As we will see in the next section (Fig.\ \ref{NA2plots}(a)), the rms thickness decreases with increasing $n_e$, leading the screening to be more effective thus letting $V_0$ smaller, if $n_e$ is varied by illumination (or by a front gate). However, it is the other way around when the back gate is used: the thickness, hence $V_0$, is expected to increase with increasing $n_e$, at variance with the experiment. Furthermore, it can be demonstrated that the change in $\epsilon_\mathrm{TF}(q)$ is too small, roughly 1\% at most, by evaluating eq.\ (\ref{Form}) using numerically calculated $\psi(z)$ that will be described in the next subsection. The observed $V_0$ vs $n_e$ requires a mechanism other than the screening for its explanation. In what follows, we neglect altogether the change in $\epsilon_\mathrm{TF}(q)$ by $n_e$.

\subsection{The conduction-band edge and the subband level}
For the range of $n_e$ and the temperature, 4.2 K, considered in the present paper, electrons occupy only the lowest subband in the confinement potential. The potential modulation seen by the electrons is the spatial variation of this subband level, $E_1$, with respect to the Fermi level. By contrast, it is the conduction-band edge $E_c$, not $E_1$, that is modified by the external devices to introduce potential modulation (a grating placed on the surface in the present study); they do not directly couple to $E_1$. The distinction is not important when the two energy levels shift in parallel, which will be approximately the case for 2DEG in a narrow quantum well. In general, however, the shift in $E_1$ does not necessarily exactly follow the shift in $E_c$. The subband level $E_1$, hence the conversion rate $dE_1/d(\delta E_c)$ from the shift $\delta E_c$ in the conduction-band edge, evaluated just above the heterointerface where wave function of the electron vanishes, to $E_1$, is dependent on the width or the profile of the confinement potential $V(z)$ in which the subband is formed. In Al$_x$Ga$_{1-x}$As/GaAs SH 2DEG, electrons themselves and the back-gate voltages are important ingredients for determining the profile of $V(z)$, as we will see below. Therefore, $dE_1/d(\delta E_c)$ varies with electron density or back-gate conditions. In what follows, we demonstrate by numerical simulation using simplified model that the observed variation of $V_0$ by the back gate or illumination is mainly attributable to the concomitant change in $dE_1/d(\delta E_c)$.

The potential well confining the electrons is given by, \cite{Ando82I,Stern84}
\begin{equation}
V(z)=V_b \Theta(-z)+(-e)\phi(z)+V_\mathrm{xc}(z).
\label{Confine}
\end{equation}
Here, we define $z$=0 as the heterointerface, and the $z$-axis points down into the substrate ($z$$<$0: Al$_x$Ga$_{1-x}$As and $z$$>$0: GaAs). $V(z)$ is comprised of the band discontinuity $V_b\Theta(-z)$ at the heterointerface ($V_b$=292 meV for $x$=0.33 at 4.2 K) and the electrostatic potential $(-e)\phi(z)$ originating from the ionized donor (Si) at doped Al$_x$Ga$_{1-x}$As layer, remnant background impurities, as well as the electrons themselves (the Hartree term). The exchange-correlation effects, the effects of electron-electron interaction beyond the Hartree approximation, is also taken into account as parameterized potential \cite{Hedin71} $V_\mathrm{xc}(z)$ after Stern and Sarma, \cite{Stern84}
\begin{eqnarray}
\lefteqn{V_\mathrm{xc}(z)} \nonumber \\
&=&\!\!\!\!\!\!-\left[1+0.7734\frac{r_s}{21}\ln\left(1+\frac{21}{r_s}\right)\right]\frac{2}{(4/9\pi)^{1/3}\pi r_s}\mathrm{Ry}^*, \nonumber \\
& &
\label{Vxc}
\end{eqnarray}
where
\begin{equation}
r_s\equiv r_s(z)=\left[\frac{4}{3}\pi{a_B^*}^3n_e|\psi(z)|^2\right]^{-1/3}
\label{rs}
\end{equation}
and $\mathrm{Ry}^*$$=$$e^2/8\pi\epsilon_0\epsilon a_B^*$ the effective Rydberg energy (5.2 meV for GaAs).
The electrostatic potential is determined by solving the Schr\"odinger equation,
\begin{equation}
-\frac{\hbar^2}{2m^*}\frac{d^2}{dz^2}\psi(z)+V(z)\psi(z)=E_1\psi(z)
\label{Shrodinger}
\end{equation}
and Posisson's equation,
\begin{equation}
\frac{d^2}{dz^2}\phi(z)=-\frac{(-e)}{\epsilon_0\epsilon}\left[n_e|\psi(z)|^2+N_A\Theta(z)\right]
\label{Poisson}
\end{equation}
self-consistently. Here we have made several types of simplifications: the difference in the effective mass and dielectric constant between the two materials is neglected and those for GaAs ($m^*$=0.067$m_e$ and $\epsilon$=13.18) are used throughout, and accordingly the image potential energy is ignored; possible atomic-scale smooth grading of the band discontinuity at the interface considered in ref.\ \citen{Stern84} is neglected; the background residual impurity $N_A$ is taken into account only at $z$$>$0. \cite{AlGaAs} These approximations are expected not to affect the present argument very much. In GaAs-based materials grown in the modern molecular beam epitaxy (MBE) machines, it is known that the residual impurity is mainly composed of carbon that is unintentionally incorporated into the crystal. In GaAs, the carbon works as an acceptor, whose density $N_A$ in the particular 2DEG wafer used in the present study is difficult to know. It can be estimated, however, to be close to 1.7$\times$10$^{20}$ m$^{-3}$ inferred from Hall measurement for nominally undoped GaAs bulk crystal grown slightly before the present 2DEG wafer was grown, using the identical MBE chamber. Therefore we use $N_A$ near this value for calculations.

The effects of the ionized donor, the background impurity, and the back gate are taken into account as suitable boundary conditions. The slope of the electrostatic potential just below the 2DEG, i.e., where $|\psi(z)|^2$ vanishes at $\langle z\rangle$$<$$z$$\ll$$z_\mathrm{depl}$ (with $\langle z\rangle$ and $z_\mathrm{depl}$ denoting the average position of the wave function and the depletion layer thickness, respectively), is determined by the electric field due to the depletion charge $n_\mathrm{depl}$ and the slope by the back gate. We parameterize the effect of the back gate by the change in the electron density $\Delta n_\mathrm{bg}$ induced by applying the back-gate voltage $V_\mathrm{bg}$; $e\Delta n_\mathrm{bg}$=$(\epsilon_0\epsilon/d_\mathrm{bg})V_\mathrm{bg}$ as defined before. Thus, by Gauss's theorem, $-d\phi/dz$$=$$(-e/\epsilon_0\epsilon)(n_\mathrm{depl}-\Delta n_\mathrm{bg})$. The slope just above the 2DEG (where $|\psi(z)|^2$ vanishes at $-d_\mathrm{sp}$$\ll$$z$$<$$0$, with $d_\mathrm{sp}$=40 nm representing the thickness of the spacer layer) results from the charges from the donor (or equivalently from the charges from 2DEG electrons in addition to the depletion charge and back-gate contribution, considering the charge neutrality), and therefore $-d\phi/dz$$=$$(-e/\epsilon_0\epsilon)(n_\mathrm{depl}+n_e-\Delta n_\mathrm{bg})$. The depletion charge and the depletion layer thickness are found by the conditions $\phi(z_\mathrm{depl})$$=$$E_g/(-e)$ and $d\phi/dz(z_\mathrm{depl})$$=$$0$, setting $\Delta n_\mathrm{bg}$$=$$0$. We obtain $n_\mathrm{depl}$$=$$(2\epsilon_0\epsilon N_AE_g/e^2)^{1/2}$ and $z_\mathrm{depl}$$=$$(2\epsilon_0\epsilon E_g/e^2N_A)^{1/2}$, respectively, the latter being the order of micro meters. As a band gap of the GaAs, we adopted $E_g$=1.52 eV, the value at 4.2 K\@.
The solution of eqs.\ (\ref{Confine}), (\ref{Shrodinger}), and (\ref{Poisson}) with the above boundary conditions are formally written as,
\begin{eqnarray}
\phi(z)\!\!\!\!\!\!&=&\!\!\!\!\!\!\left.\frac{(-e)}{\epsilon_0\epsilon}\right\{(n_\mathrm{depl}-\Delta n_\mathrm{bg})z \nonumber \\
& &\left.-\Theta(z)\frac{1}{2}N_Az^2+n_e\left[I(z)-I(z_\mathrm{depl})\right]\right\},
\label{phiz}
\end{eqnarray}
with
\begin{equation}
I(z)\equiv z-\int_{-\infty}^z\!\!\!\!dz^\prime\int_{-\infty}^{z^\prime}\!\!\!\!dz^{\prime\prime}|\psi(z^{\prime\prime})|^2.
\label{Iz}
\end{equation}

\begin{figure}[tb]
\includegraphics[bbllx=20,bblly=50,bburx=420,bbury=800,width=8.5cm]{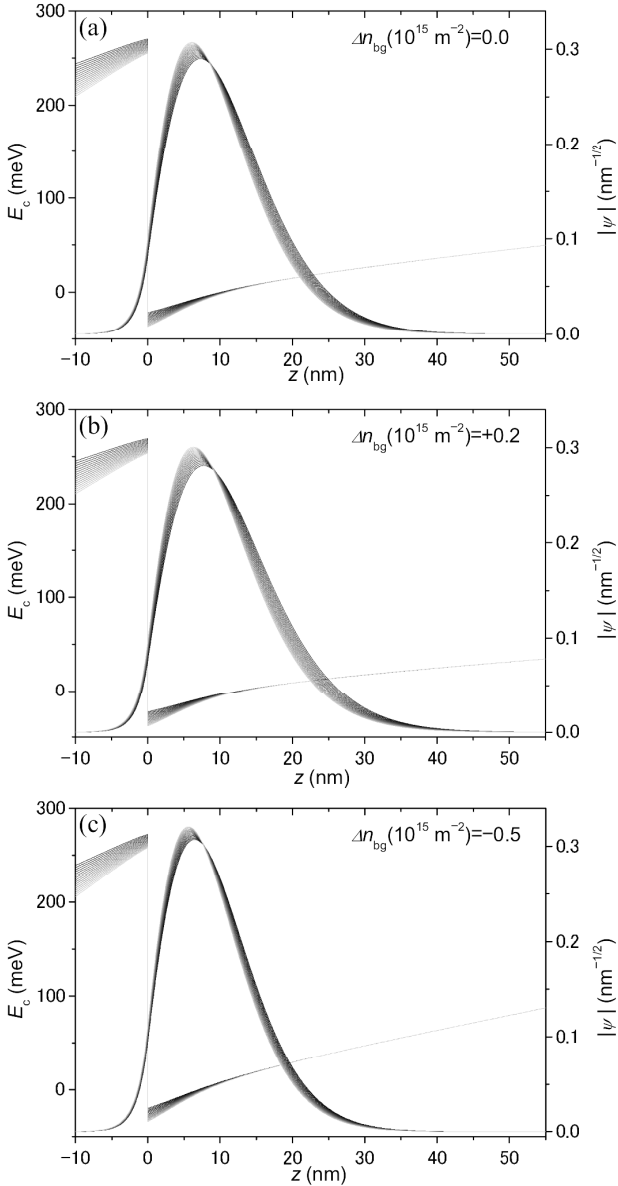}
\caption{Numerically calculated conduction-band edge $E_c$ and the wave function $\psi$, for $N_A$=2$\times$10$^{20}$ m$^{-3}$. The heterointerface resides at $z$=0. Calculations were done for the values of $n_e$(in 10$^{15}$ m$^{-2}$) from 1.5 to 3.0 by the increment of 0.1 each, and displayed with progressively brighter grayscale. The back gate was set (a) neutral ($\Delta n_\mathrm{bg}$(in 10$^{15}$ m$^{-2}$)=$0.0$) (b) positive ($+0.2$), and (c) negative ($-0.5$).}
\label{Ecwf}
\end{figure}
\begin{figure}[tb]
\includegraphics[bbllx=20,bblly=50,bburx=420,bbury=800,width=8.5cm]{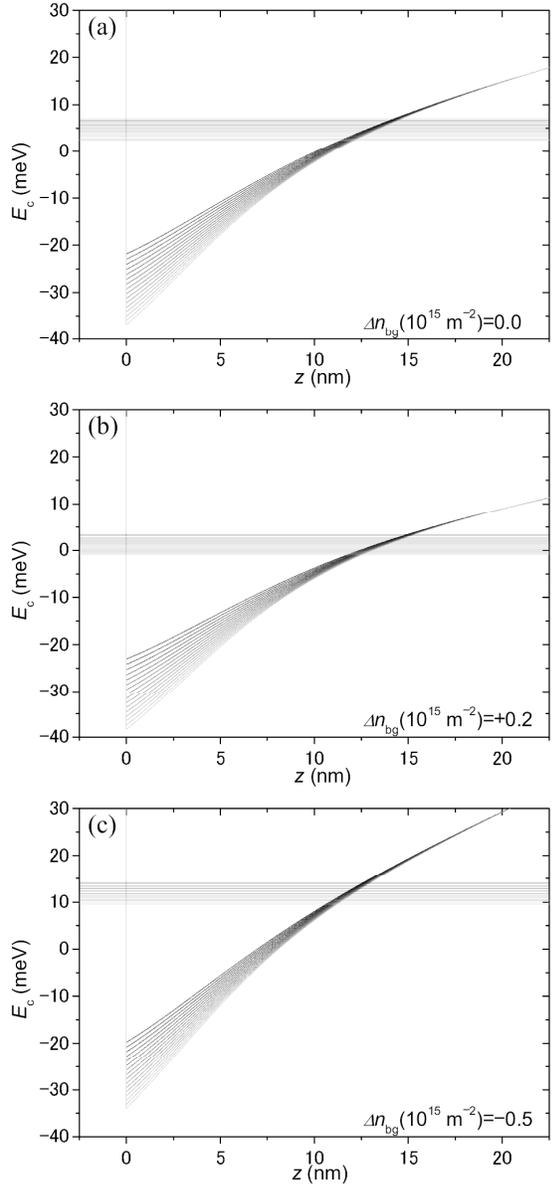}
\caption{The conduction-band edge $E_c$ near the heterointerface and the lowest-subband level $E_1$. The values of $N_A$, $n_e$, and $\Delta n_\mathrm{bg}$, and the grayscale are the same as in Fig.\ \ref{Ecwf}}
\label{EcE1}
\end{figure}

\begin{figure}[tb]
\includegraphics[bbllx=20,bblly=30,bburx=420,bbury=800,width=8.5cm]{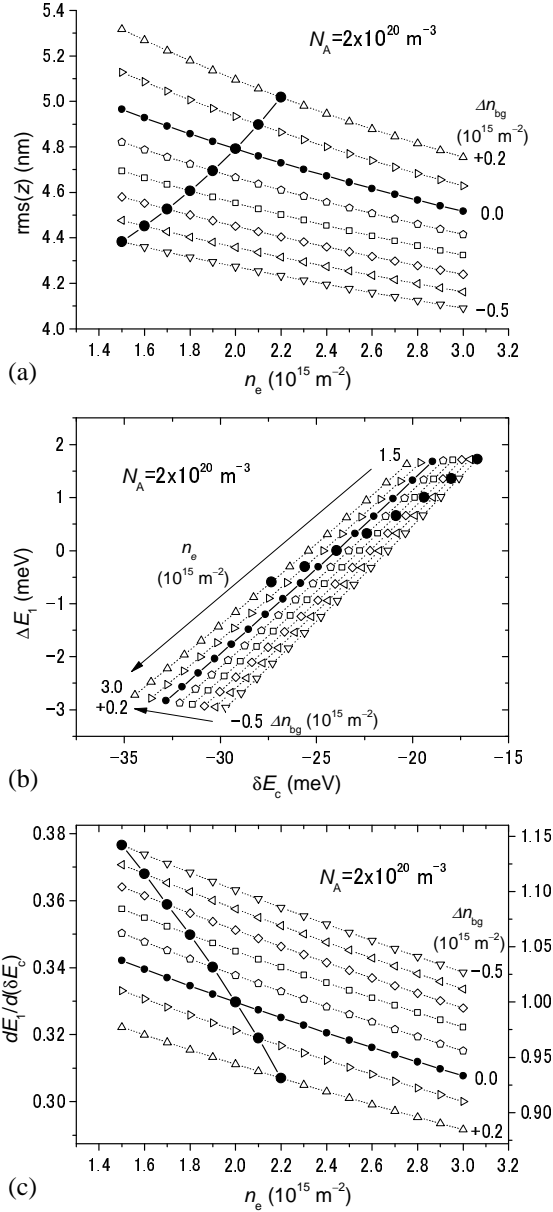}
\caption{(a) rms thickness of the wave function vs $n_e$, (b) the shift $\Delta E_1$ of the lowest subband level $E_1$ (measured from the value at $n_e$=2.0$\times$10$^{15}$ m$^{-2}$) vs the shift $\delta E_c$ of the conduction-band edge just above the heterointerface, (c) the rate of the change of $E_1$ in response to the change in $\delta E_c$, for different settings of the back gate ($\Delta n_\mathrm{bg}$ from $-0.5$ to $+0.2$$\times$10$^{15}$ m$^{-2}$ at the interval of $0.1$$\times$10$^{15}$ m$^{-2}$, each plotted by differently shaped symbols), and $N_A$=2$\times$10$^{20}$ m$^{-3}$. The large (small) solid circles designate how the values evolve when $n_e$ is varied by the back gate (by illumination). The right axis in (c) shows the ratio with the value at $n_e$=2.0$\times$10$^{15}$ m$^{-2}$ and $\Delta n_\mathrm{bg}$=0.0$\times$10$^{15}$ m$^{-2}$.}
\label{NA2plots}
\end{figure}

Numerical solutions of the wave function $\psi(z)$, and the $z$ dependence of the conduction-band edge $E_c$, which is no other than the confinement potential $V(z)$, is plotted in Fig.\ \ref{Ecwf}. Figure \ref{Ecwf}(a) to (c) correspond to different settings of the back-gate voltages $\Delta n_\mathrm{bg}$$=$$(\epsilon_0\epsilon/ed_\mathrm{bg})V_\mathrm{bg}$. For each $\Delta n_\mathrm{bg}$, calculations are repeated for different $n_e$'s ranging from 1.5 to 3.0$\times$10$^{15}$ m$^{-2}$ with an increment of 0.1$\times$10$^{15}$ m$^{-2}$ each. As a matter of course, the confinement potential, hence the wave function, becomes thicker (thinner) for a positive (negative) back-gate voltage. For a fixed $\Delta n_\mathrm{bg}$, the width of $\psi(z)$ decreases with increasing $n_e$, owing to the last term in eq.\ (\ref{phiz}). The change in the thickness is more conspicuous for larger (more positive) $\Delta n_\mathrm{bg}$. This is more clearly illustrated in Fig.\ \ref{NA2plots}(a), which plots rms thickness of the wave function versus $n_e$ for varying settings of $\Delta n_\mathrm{bg}$ ranging from $-$0.5 to $+$0.2 with an interval of 0.1 (in 10$^{15}$ m$^{-2}$). Figure \ref{EcE1} shows close-up of $E_c$ in the proximity of the heterointerface along with the lowest subband level $E_1$. The conduction-band edge $E_c$ at the heterointerface, $z$=0, displays downward shift with the increase of $n_e$. Accompanying this deepening of the bottom of the confinement potential, the subband level $E_1$ is also displaced downward but with smaller decrement. Close inspection of Fig.\ \ref{EcE1}(a)--(c) reveals that the downward shift in $E_c$ by $n_e$ is larger (smaller) for more positive (negative) back-gate voltages, while the extent of shift in $E_1$ remains virtually unchanged. $E_c$ at the heterointerface approximately represents the $E_c$ just above 2DEG, the $E_c$ that we will be looking at (to be denoted as $\delta E_c$), apart from minor modification caused by not including the contribution from the tail of the wave function in the Al$_{x}$Ga$_{1-x}$As barrier layer and including the contribution from $V_\mathrm{xc}(0)$. More quantitative account is given in Fig.\ \ref{NA2plots}(b).
\begin{figure}[tb]
\includegraphics[bbllx=20,bblly=80,bburx=600,bbury=420,width=8.5cm]{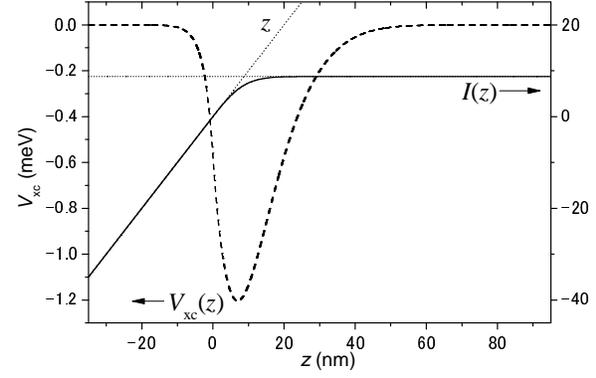}
\caption{$V_\mathrm{xc}(z)$ (dashed curve, left axis) and $I(z)$ (solid curve, right axis) calculated for $N_A$=2$\times$10$^{20}$ m$^{-3}$, $n_e$=2.0$\times$10$^{15}$ m$^{-2}$, and $\Delta n_\mathrm{bg}$=0.0$\times$10$^{15}$ m$^{-2}$.}
\label{VxcIz}
\end{figure}

Before explaining Fig.\ \ref{NA2plots} further, we look into more detail of the behavior of the last term in eq.\ (\ref{phiz}), which governs the $n_e$ dependences we have just pointed out. The function $I(z)$ appearing in the term tends to $z$ for $z$$\ll$0, and to a constant value dependent on the distribution of $\psi(z)$ for large enough $z$($\gg$$\langle z\rangle$), as exemplified in Fig.\ \ref{VxcIz} for particular values of $N_A$, $n_e$, and $\Delta n_\mathrm{bg}$. It can easily be perceived from the definition eq.\ (\ref{Iz}) that the constant value at $z$$\gg$$\langle z\rangle$, $I(\infty)$[$\simeq$$I(z_\mathrm{depl})$ since $z_\mathrm{depl}$$\gg$$\langle z\rangle$], increases with increasing width of $\psi(z)$. Physically, this reflects the property of the solution of the Poisson's equation that the potential change across a space-charge layer with a given amount of charge is larger when the charge is spatially spread over more widely. These features are transparent for Fang-Howard wave function, which allows analytic evaluation as $I(z)$$=$$z-\Theta(z)[z-3/b+\exp(-bz)(3+2bz+b^2z^2/2)/b]$ and $I(z_\mathrm{depl})$$\simeq$$I(\infty)$$=$$3/b$.

The net shift in $E_c$ across the 2DEG (from $z$$<$0 to $z$$>$0) brought about by electrons themselves is given by
\begin{equation}
\delta E_c=-\frac{e^2}{\epsilon_0\epsilon}n_eI(z_\mathrm{depl}),
\label{deltaEc}
\end{equation}
which includes the contribution from the tail of $\psi(z)$ in the barrier layer. Although $V_\mathrm{xc}(z)$ also depends on $n_e$, it does not contribute to the net shift since it vanishes where $\psi(z)$ vanishes (see eqs.\ (\ref{Vxc}), (\ref{rs}) and an example in Fig.\ \ref{VxcIz}). Note that $I(z_\mathrm{depl})$=0 for an ideal zero-thickness 2DEG $|\psi(z)|^2$$=$$\delta(z)$, therefore eq.\ (\ref{deltaEc}) is the effect resulting exclusively from the finite thickness.

 It is worth pointing out that the last term in eq.\ (\ref{phiz}) is the only ``plastic'' term in $\phi(z)$ that can be modified externally from the front surface. All the other terms are fixed once the sample quality ($N_A$) and the back-gate setting ($\Delta n_\mathrm{bg}$) are fixed. An attempt to introduce potential modulation from the surface, e.g., by the grating induces perturbing spatial variation in $E_c$ above the 2DEG, namely, in $\delta E_c$. Following the change in $\delta E_c(\textbf{r})$ at the position $\textbf{r}$ in the $x$-$y$ plane, $n_e(\textbf{r})$, $\psi(\textbf{r},z)$, and hence $E_1(\textbf{r})$ are also slightly altered to fulfill eqs.\ (\ref{Confine}), (\ref{Shrodinger}), (\ref{Poisson}) and (\ref{deltaEc}) with the modified $\delta E_c$, thereby resulting in the spatial variation in $E_1$. When the modulation amplitude is small enough, as is the case in the present study, the conversion ratio from $\delta E_c$ to $E_1$ can be evaluated by the derivative $dE_1/d(\delta E_c)$ at the original $n_e$. Further, with a plausible assumption that the amplitude of the variation in $\delta E_c$ to be introduced from the surface does not depend on properties at and below 2DEG, i.e., on $n_e$, $n_\mathrm{depl}$, and $\Delta n_{bg}$, the amplitude $V_0$ seen by electrons will be proportional to $dE_1/d(\delta E_c)$.

In Fig.\ \ref{NA2plots}(b), $E_1$ is plotted against $\delta E_c$ given by eq.\ (\ref{deltaEc}). (To plot all of them within a single frame, $E_1$'s are negatively offset by the value at $n_e$=2.0$\times$10$^{15}$ m$^{-2}$ for each $\Delta n_\mathrm{bg}$, hence the notation $\Delta E_1$.) The plots are for the same sets of $n_e$ and $\Delta n_\mathrm{bg}$ as in (a). Applying external perturbation to introduce modulation corresponds to slightly shifting the point ($\delta E_c$,$\Delta E_1$) along the curve of constant $\Delta n_\mathrm{bg}$, because the perturbation is assumed not to affect $\Delta n_\mathrm{bg}$. Therefore the slope of the constant $\Delta n_\mathrm{bg}$ curve is the $dE_1/d(\delta E_c)$ to be considered here. It can be seen in Fig.\ \ref{NA2plots}(b) and clearer in Fig.\ \ref{NA2plots}(c), which plots \cite{Poly} $dE_1/d(\delta E_c)$ versus $n_e$, that the slope has a trend of becoming smaller with increasing $n_e$ if $\Delta n_\mathrm{bg}$ is kept constant, and also with increasing (more positive) $\Delta n_\mathrm{bg}$. It is to this trend that we ascribe the observed behavior of $V_0$. The decrease of $dE_1/d(\delta E_c)$ by increasing $\Delta n_\mathrm{bg}$ is readily interpretable as an effect of the thickness of the 2DEG: the negative shift of $\delta E_c$ with $n_e$ is more rapid for thicker 2DEG while the decrease in $E_1$ does not change very much (see Fig.\ \ref{EcE1} or Fig.\ \ref{NA2plots}(b)), and therefore $dE_1/d(\delta E_c)$ is smaller for thicker 2DEG\@. The decrease of $dE_1/d(\delta E_c)$ with increasing $n_e$ for constant $\Delta n_\mathrm{bg}$ originates from more subtle competition between the shift in $\delta E_c$ and $E_1$: both shift downward with increasing $n_e$, and the rate of the shift decreases with increasing $n_e$ whose rate (the rate of ``the rate of shift'') decelerating slightly more rapidly for $\delta E_c$, leaving the rate of change for $\delta E_c$ relatively larger than that of $E_1$.

In Fig.\ \ref{NA2plots}, the results of the calculations are plotted for all possible sets of $n_e$ and $\Delta n_\mathrm{bg}$. In the experiment, $n_e$ was varied either by the back gate alone, namely $n_e$$=$$n_{e0}+\Delta n_\mathrm{bg}$ with $n_{e0}$ fixed, or by illumination with the back-gate voltage fixed to zero $\Delta n_\mathrm{bg}$=0. The plotted points corresponding to the two experimental modes are highlighted by large and small solid circles, respectively, with $n_{e0}$ set to 2.0$\times$10$^{15}$ m$^{-2}$. Figure \ref{NA2plots}(a) shows that rms thickness increases with increasing $n_e$ when varied by the back gate, while the trend is reversed when $n_e$ is varied by illumination. As demonstrated in Fig.\ \ref{NA2plots}(c), on the other hand, $dE_1/d(\delta E_c)$ decreases for both methods with the rate of the change much larger when $n_e$ is driven by the back gate. Thus, $dE_1/d(\delta E_c)$ behaves in qualitatively the same manner as $V_0$ under the variation of $n_e$ by the back gate and by illumination. 

\begin{figure}[tb]
\includegraphics[bbllx=20,bblly=60,bburx=600,bbury=420,width=8.5cm]{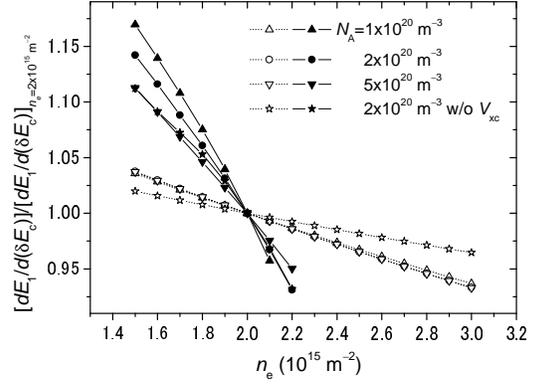}
\caption{The rate of change $dE_1/d(\delta E_c)$ (normalized by the value at $n_e$=2.0$\times$10$^{15}$ m$^{-2}$ and $\Delta n_\mathrm{bg}$=0.0$\times$10$^{15}$ m$^{-2}$) plotted as a function of $n_e$ for different values of $N_A$. Solid and open symbols correspond to varying $n_e$ by the back gate and by illumination, respectively. Stars represent the calculation without $V_\mathrm{xc}(z)$ in eq.\ (\ref{Confine}).}
\label{NAdep}
\end{figure}

To see how $N_A$ affects the problem, we calculate $dE_1/d(\delta E_c)$ for two other values of $N_A$ through the same procedure as was done for Fig.\ \ref{NA2plots}(c), and plot in Fig.\ \ref{NAdep} the $dE_1/d(\delta E_c)$ normalized by its value at $n_e$=2.0$\times$10$^{15}$ m$^{-2}$. The figure shows that the behavior does not depend very much on $N_A$ when $n_e$ is varied by illumination, while the value of $N_A$ has strong impact on the rate of decrease when the back gate is in use; the decrease is more rapid for smaller $N_A$, i.e., for cleaner 2DEG\@. This is readily interpretable in terms of its effect on the confinement potential. As manifest from eq.\ (\ref{phiz}), the confinement potential has the same profile as long as $n_\mathrm{depl}^\prime$=$n_\mathrm{depl}-\Delta n_\mathrm{bg}$, which should be positive for the confinement to be operative, remains unchanged. Small $N_A$ hence small $n_\mathrm{depl}$ intensifies the relative importance of the variation of the $\Delta n_\mathrm{bg}$. Large solid circles in Fig.\ \ref{NA2plots}(c) shows that the rate of the change of $dE_1/d(\delta E_c)$ versus $n_e$, varied by the back gate, increases with increasing (more positive) $\Delta n_\mathrm{bg}$, namely, with increasing thickness of the 2DEG\@. Since decreasing $n_\mathrm{depl}$ is equivalent to increasing $\Delta n_\mathrm{bg}$, it also adds to the change rate. The qualitative behavior of $dE_1/d(\delta E_c)$ under the variation of $n_e$, $N_A$, and $\Delta n_\mathrm{bg}$ is reproduced in a simple analytic calculation using Fang-Howard wave function (see Appendix).

For quantitative comparison with the experiment, the plots in Fig.\ \ref{NAdep} for $N_A$=1.0 and 2.0$\times$10$^{20}$ m$^{-3}$ are replotted in Fig.\ \ref{V0plot}(b). \cite{Poly2} As mentioned before, the rationale of comparing $V_0$ with $dE_1/d(\delta E_c)$ is given by the assumption that the amplitude of the periodic modulation in $\delta E_c$ is determined solely by the grating on the surface and independent of $n_e$, $N_A$, and $\Delta n_\mathrm{bg}$; from the assumption, the proportionality of $V_0$ and $dE_1/d(\delta E_c)$ is expected to result. (Note, however, that it is only the proportionality that can be compared with experiment, since we currently do not have a method to directly measure the amplitude of the modulation of $\delta E_c$.) Fig.\ \ref{V0plot}(b) demonstrates that the behavior of $V_0$ under back gate and illumination is explained by that of $dE_1/d(\delta E_c)$ fairly well. The agreement is better for $N_A$=1.0$\times$10$^{20}$ m$^{-3}$, implying that the density of residual impurity in our 2DEG wafer may be closer to this value (or smaller). However, since we have used simplified model for the confinement potential, and have neglected many complications, e.g., possible screening by the Si-doped layer, \cite{Davies94} it seems going too far to use this comparison for quantitatively accurate determination of $N_A$.

Finally, we discuss the effect of the exchange and correlation term $V_\mathrm{xc}(z)$. In Fig.\ \ref{NAdep} we also plot the result of calculation deliberately omitting the term $V_\mathrm{xc}(z)$ from eq.\ (\ref{Confine}). The plot, along with Fig.\ \ref{V0plot}(b), illustrates that experiments are better described with inclusion of $V_\mathrm{xc}(z)$. The predominant effect of $V_\mathrm{xc}(z)$ is twofold: it reduces the thickness of the 2DEG hence the downward shift of $\delta E_c$, and also reduces $E_1$ by partially alleviating the penalty in energy due to the Hartree term. It turns out by examining the process of the calculation that the decrease in $E_1$ by $V_\mathrm{xc}(z)$ dominates the difference with and without the term. The effect is more pronounced for larger $n_e$, letting the decrease of $dE_1/d(\delta E_c)$ with $n_e$ more rapid and closer to the experiment. This underlines the importance of the exchange-correlation effect, albeit the smallness of its magnitude, in understanding the behavior of the potential modulation in the 2DEG\@. 

\section{Conclusions}
\label{Conclusions}
We have reported our experimental result that the amplitude $V_0$ of potential modulation decreases with increasing electron density $n_e$, varied either by the back gate or by illumination. Contrary to the intuition, the decrease is much more rapid when the back gate is employed. We have ascribed the result to the modification in the conversion rate $dE_1/d(\delta E_c)$ for the perturbing modulation with its source at the surface (or more generally, above the 2DEG) to be transmitted to the subband level in the confinement potential $V(z)$. The origin of the substantial change in $dE_1/d(\delta E_c)$ can be traced back to the high sensitivity, particularly notable in a SH 2DEG, of the profile of $V(z)$ hence of the envelope function $\psi(z)$ to the change in the electron density or the back-gate setting. Therefore we expect much smaller effect in a 2DEG formed in a single quantum well. The present result provides a prescription for making the modulation amplitude large in SH 2DEG; that is, it is advantageous to make the 2DEG as thin as possible. This is especially important for a small period ULSL for which the amplitude becomes inevitably small.

Although we have in the present paper confined our interest to ULSL, for which the measurement of the modulation amplitude $V_0$ is possible, the decrease of $V_0$ with increasing $n_e$ will also take place likewise in other low-dimensional electron systems based on SH 2DEG with potential modulation. Moreover, even in plain SH 2DEGs without any artificial modulation, the random potential landscape seen by electrons, since its main source is ionized donors located \textit{above} the 2DEG, will diminish its amplitude with increasing $n_e$. The effect should be born in mind in interpreting the experiment with variation of $n_e$, especially when $n_e$ is varied by the back gate. \cite{Soibel97,Endo04EP,Goldman95}

\section*{Acknowledgment}
This work was supported by Grant-in-Aid for Scientific Research (C) (15540305) and (A) (13304025) and Grant-in-Aid for COE Research (12CE2004) from Ministry of Education, Culture, Sports, Science and Technology.

\appendix
\section{Fang-Howard approximation}
The Fang-Howard approximation, although not quantitatively quite accurate, provides analytic formula, which will be useful for the qualitative understanding of the phenomenon. Here, we calculate $dE_1/d(\delta E_c)$ using Fang-Howard wave function $\psi_\mathrm{FH}(z)$. The lowest subband energy is given by the sum of the expectation values of the kinetic energy and the potential energy, $E_1$=$\langle T\rangle$+$\langle V\rangle$, with $\langle T\rangle$=$\langle \psi_\mathrm{FH}(z)|(-\hbar^2/2m^*)(d^2/dz^2)|\psi_\mathrm{FH}(z)\rangle$=$\hbar^2b^2/8m^*$. In evaluating potential energy, we neglect $V_\mathrm{xc}(z)$ for simplicity. Noting that $\psi_\mathrm{FH}(z)$ vanishes at $z$$<$0, we obtain, using eq.\ (\ref{phiz}),
\begin{eqnarray}
\!\!\!\!\langle V\rangle\!\!\!\!&=&\!\!\!\!\langle\psi_\mathrm{FH}(z)|(-e)\phi(z)|\psi_\mathrm{FH}(z)\rangle \nonumber \\
&=&\!\!\!\!\frac{3e^2}{\epsilon_0\epsilon}\left(\frac{n_\mathrm{depl}-\Delta n_\mathrm{bg}-5n_e/16}{b}-\frac{2N_A}{b^2}\right),
\end{eqnarray}
thus,
\begin{eqnarray}
\!\!\!\!E_1\!\!\!\!&=&\!\!\!\!\mathrm{Ry}^*\left[\frac{1}{4}{a_B^*}^2b^2\right. \nonumber \\
&&\!\!\!\!\left.+24\pi a_B^*\left(\frac{n_\mathrm{depl}^\prime-5n_e/16}{b}-\frac{2N_A}{b^2}\right)\right], \nonumber\\
&&
\label{FHE1}
\end{eqnarray}
with $n_\mathrm{depl}^\prime$$\equiv$$n_\mathrm{depl}-\Delta n_\mathrm{bg}$. As mentioned before,
\begin{equation}
\delta E_c=-\frac{e^2}{\epsilon_0\epsilon}n_e\frac{3}{b}=-24\pi \mathrm{Ry}^*a_B^*n_e\frac{1}{b}.
\end{equation}
Making use of the relation \cite{AndoR82,variational} $b$=$[48\pi(n_\mathrm{depl}^\prime+11n_e/32)/a_B^*]^{1/3}$, and being reminded that $N_A$ hence $n_\mathrm{depl}$, and $\Delta n_{bg}$ are fixed, $b$ can be made the only free parameter by eliminating $n_e$, and therefore,
\begin{eqnarray}
\!\!\!\!\frac{dE_1}{d(\delta E_c)}\!\!\!\!&=&\!\!\!\!\frac{dE_1/db}{d(\delta E_c)/db} \nonumber \\
&=&\!\!\!\!\frac{(3a_B^*/16\pi)b^3+21n_\mathrm{depl}^\prime-44N_A/b}{(4a_B^*/3\pi)b^3+32n_\mathrm{depl}^\prime} \nonumber\\
&=&\!\!\!\!\frac{5}{16}\left[\frac{n_\mathrm{depl}^\prime+(33/320)n_e-(22/15)N_A/b}{n_\mathrm{depl}^\prime+(11/48)n_e}\right]. \nonumber \\
&&
\label{FHdE1dEc}
\end{eqnarray}
In the last equality, the parameter $b$ is replaced back again to $n_e$, except for the term including $N_A$, which is negligibly (roughly three orders of magnitude) small compared with other terms. It is easy to verify from eq.\ (\ref{FHdE1dEc}) that $dE_1/d(\delta E_c)$ decreases with increasing $n_e$ when (i) $n_{e0}$$\equiv$$n_e-\Delta n_\mathrm{bg}$ is fixed (simulating the back-gate control) and (ii) $n_\mathrm{depl}^\prime$ is kept constant (simulating the illumination), and the rate of change is larger for the former.


\begin{thebibliography}{99}
\bibitem{density} The electron density slightly ($\leq$ 5 \%) decreases on depositing the back gate even when $V_\mathrm{bg}$=0. This is attributable to the large work function of gold, which effectively work as natively negative voltage.
\bibitem{harmonics} Higher order harmonics are neglected throughout this paper. A quantitative account of higher-harmonic contents will be presented elsewhere.
\bibitem{AlGaAs} In reality, because it contains more reactive element Al, Al$_x$Ga$_{1-x}$As is expected to include impurities at higher concentration than GaAs. However, $N_A$ is important only to determine $n_\mathrm{depl}$ by the boundary condition deep inside the substrate. The direct effect of $N_A$ near the interface is negligibly small. Therefore neglecting the background impurities in the Al$_x$Ga$_{1-x}$As layer is legitimate.
\bibitem{Poly} The differentiation is actually performed on a quadratic polynomial fitted to the discrete points in Fig.\ \ref{NA2plots}(b).
\bibitem{Poly2} Strictly, quadratic polynomials fitted to the discrete points in Fig.\ \ref{NAdep} are plotted to make the comparison with the experimental plots easy to see.
\bibitem{variational} The parameter $b$ is optimized by minimizing the energy slightly different from $E_1$ given in eq.\ (\ref{FHE1}) in the calculation of $\langle V\rangle$. Namely, (i) the Hartree term, resulting from the term containing $n_eI(z)$ in eq.\ (\ref{phiz}), is halved to avoid double count of the electron-electron interaction, \cite{AndoR82} (ii) the term from the final term including $n_eI(z_\mathrm{depl})$ in eq.\ (\ref{phiz}), which was necessary to conform the bondary condition at $z$=$z_\mathrm{depl}$$\gg$0, is omitted, and (iii) the term including $N_A$ is neglected because of its smallness.
\end{thebibliography}
\end{document}